\documentclass[
preprint,
showpacs,
amsmath,amssymb,
aps,prf,longbibliography
]{revtex4-1}
\usepackage{amssymb}
\usepackage{amsmath}
\usepackage{booktabs}
\usepackage{tabularx}
\usepackage{braket}
\usepackage{float}
\usepackage{graphicx}
\usepackage{dcolumn}
\usepackage{bm}
\usepackage{subfigure}
\usepackage{color}

\usepackage{epstopdf, epsfig}
\usepackage{blindtext}
\usepackage{longtable}
\usepackage{amsmath}
\usepackage{mathtools}
\usepackage{enumitem}
\usepackage[T1]{fontenc}
\usepackage{mathptmx}
\usepackage{overpic}

\usepackage[colorlinks,citecolor = blue, linkcolor=blue,hyperindex,CJKbookmarks]{hyperref}

\begin{document}

\title{Universal properties of 
penetrative turbulent Rayleigh--B\'enard convection in cold water near $4^\circ\rm{C}$}

\author{Qi Wang$^{1,2}$}
\author{Philipp Reiter$^{3}$}
\author{Detlef Lohse$^{1,3}$}\email{d.lohse@utwente.nl}
\author{Olga Shishkina$^{3}$}\email{Olga.Shishkina@ds.mpg.de}
\affiliation{$^1$Physics of Fluids Group and Max Planck Center for Complex Fluid Dynamics, MESA+ Institute and J. M. Burgers Centre for Fluid Dynamics, University of Twente, P.O. Box 217, 7500AE Enschede, The Netherlands\\
$^2$Department of Modern Mechanics, University of Science and Technology of China, Hefei 230027, China\\
$^3$Max Planck Institute for Dynamics and Self-Organization, 37077 G\"ottingen, Germany}

\date{\today}

\begin{abstract}
Penetrative turbulence, which occurs in a convectively unstable fluid layer and penetrates into an adjacent, originally stably stratified layer,  is numerically and theoretically analyzed. 
We chose the 
most relevant  example, namely
  thermally driven flow of water with a temperature 
  around $T_m\approx 4^\circ\rm{C}$, where it has its density maximum. We pick the Rayleigh-B\'enard geometry 
  with the  bottom  plate temperature $T_b >  4^\circ\rm{C}$
   and the top plate temperature $T_t \le  4^\circ\rm{C}$. Next to the overall thermal driving strength set by the temperature 
   difference $\Delta = T_b - T_t$ (the Rayleigh
   number $Ra$ in dimensionless form), the crucial new control parameter as compared to standard Rayleigh-B\'enard convection
   is the density inversion parameter $\theta_m \equiv (T_m - T_t ) / \Delta$. The crucial response parameters are the relative mean 
   mid-height temperature $\theta_c$ and the overall heat transfer (i.e., the Nusselt number $Nu$). 
We theoretically derive the universal (i.e., $Ra$-independent) dependence  $\theta_c (\theta_m) =(1+\theta_m^2)/2$, which holds 
for $\theta_m$ below a  $Ra$-dependent  critical value, beyond which  $\theta_c (\theta_m)$ 
 sharply decreases and drops down to  $\theta_c=1/2$ at $\theta_m=\theta_{m,c}$. Our direct
  numerical simulations with 
$Ra$ up to $10^{10}$ are consistent with these results. 
The critical density inversion parameter 
$\theta_{m,c}$ can be precisely predicted by a
 linear stability analysis. The heat flux $Nu(\theta_m)$ monotonically decreases with increasing $\theta_m$ and we can theoretically derive a universal relation 
for the relative heat flux
 $Nu(\theta_m)/Nu(0)$. 
 Finally, we numerically identify and discuss rare transitions between different  turbulent flow states for large $\theta_m$.
\end{abstract}

\maketitle

Turbulent Rayleigh--B\'enard convection (RBC) \citep{ahlers2009heat,lohse2010small,chilla2012new}, 
which occurrs in a fluid layer due to a temperature difference at its bottom and top surfaces, is 
the paradigmatic model system to study thermally driven turbulence.
Usually it is considered within the Oberbeck--Boussinesq (OB) approximation, 
in which all fluid properties are assumed to be constant, apart from the density in the buoyancy term of the momentum equation, where it is assumed to be linearly dependent on the temperature.
However, in many cases in nature and technology,
the density of many fluids is strongly non-linear and even non-monotonic with the temperature, which significantly influences the flow patterns and the heat transport properties in the system. 
The most famous and relevant example is water, for which  the density 
is maximal at $T_m\approx4^\circ\rm{C}$.
This density maximum has a drastic
  influence on many natural phenomena like the freezing of lakes and estuaries, and the survival of fauna in shallow waters in winter \citep{dumore1953heat,carey1980buoyancy,carey1981visualization,wang2020growth}.  
  However, also here RBC can be used again as paradigmatic model system, but now 
 correspondingly  RBC of cold water near $T_m$.  
 This is an example of 
  so-called penetrative convection \citep{veronis1963penetrative,musman1968penetrative}, where 
  convection in a thermally unstable layer penetrates into the adjacent stable layers. Penetrative convection is 
  not only important for water around $4^\circ$C and freezing lakes etc., but also  in many astrophysical settings, 
for example, in the tachocline of the Sun \citep{dintrans2005spectrum}, and possibly in the liquid core of the Earth \citep{buffett2014geomagnetic} and in the Jupiter's atmosphere \citep{zhang1996penetrative}.

The control parameters in this problem are 
the Prandtl number $Pr\equiv \nu/\kappa$, 
where $\nu$ is the kinematic viscosity and $\kappa$  the thermal diffusivity of the fluid,
the aspect ratio $\Gamma \equiv W/H$  as  ratio of the width $W$ and the height $H$ of the domain, 
 and 
the Rayleigh number $Ra\equiv g\alpha \Delta^q H^3/\nu \kappa$, where $H$ is the height of the cell and $g$ the gravitational
acceleration. Note that the nonlinear dependence of $Ra$ on the temperature difference $\Delta  \equiv T_b-T_t$ between the
hot bottom plate temperate $T_b > T_m$ and the cold top plate temperature $T_t \le T_m $ reflects 
  the  nonlinear relationship between the density and the temperature around the density maximum, namely
  $\rho=\rho_m(1-\alpha{|T-T_m|}^q)$ \citep{gebhart1977new}, where $\rho_m\approx1000kg/m^3$ is the maximum density at the temperature $T_m\approx4^\circ\rm{C}$. Here we take the measured material properties for water, namely 
$q=1.895$ and $\alpha=9.30\times10^{-6}(K)^{-q}$ as 
isobaric thermal expansion coefficient. 
Apart from the density, all other material properties of the fluid are assumed to be constant.
The crucial new control parameter in penetrative convection as compared to standard RBC is 
 the density inversion parameter $\theta_m\equiv(T_m-T_t)/\Delta$. 
 We consider $0 \le \theta_m < 1$, with 
 the limiting case 
 $\theta_m = 0$ (or $T_t= T_m$) without a stably stratified region and thus 
 being closest to standard RBC and the other limiting case 
 $\theta_m = 1$ (or $T_b = T_m$) being fully stably stratified.

 The response parameters of the system are the Nusselt number $Nu\equiv{QH}/{(k\Delta)}$
  and the Reynolds number $Re\equiv{UH}/\nu$,
which indicate the non-dimensional heat transport and flow strength in the system, respectively. 
Here $Q$ is the heat flux crossing the system,
$k$ the thermal conductivity,
 and $U\equiv{\left<\boldsymbol{u}^2\right>^{1/2}_{V,\,t}}$ the time and volume averaged root-mean-square velocity of the flow.
The central new response parameter in penetrative convection 
as compared to standard RBC is the mean temperature $T_c$ at mid height, or,
when non-dimensionalized, $\theta_c = (T_c - T_t )/\Delta$. Whereas in standard RBC $\theta_c = 1/2$ for symmetry reasons,
here in penetrative convection  $1/2 \le \theta_c \le 1$,  reflecting that the hot fluid from the bottom unstably stratified region 
can more easily reach the center than the cold fluid from top stably stratified region.

Early studies of penetrative RBC  focused on the stability of the flows near the onset of convection \citep{veronis1963penetrative, musman1968penetrative, moore1973nonlinear, walden1981, large2014penetrative}. Recently, turbulent RBC of cold water near $T_m$ also attracted attention \citep{lecoanet2015numerical,wang2019penetrative,wang2020growth}. In Ref.~\cite{lecoanet2015numerical}, 
the internal gravity wave excitation by convection was studied using two-dimensional (2D) direct numerical simulations (DNS). 
In Ref.~\cite{wang2019penetrative}, convection in cold water was investigated by means of 2D and three-dimensional (3D) DNS. It was shown that for $0\le\theta_m\lesssim 0.9$, the mean central temperature $\theta_c$, the normalized Nusselt number $Nu(\theta_m)/Nu(0)$, and the normalized Reynolds number $Re(\theta)/Re(0)$ are almost independent of $Ra$ and thus to be claimed to be universal.

However, in this study we  
will show that when extending the range of the control parameter $\theta_m$ towards $1$ (i.e., towards the fully stably stratified case),
the center temperature $\theta_c$ strongly depends 
 on $Ra$.   
Moreover, we propose a model for the dependence of 
this key response parameter $\theta_c$ on the density inversion parameter $\theta_m$ 
 and demonstrate its excellent agreement with the DNS data.
Also the dependence of the other central response parameter, 
namely the Nusselt number,   on $\theta_m$ is theoretically explained.

  \begin{figure*}
	\centering
	\begin{overpic}[width=0.96\textwidth]{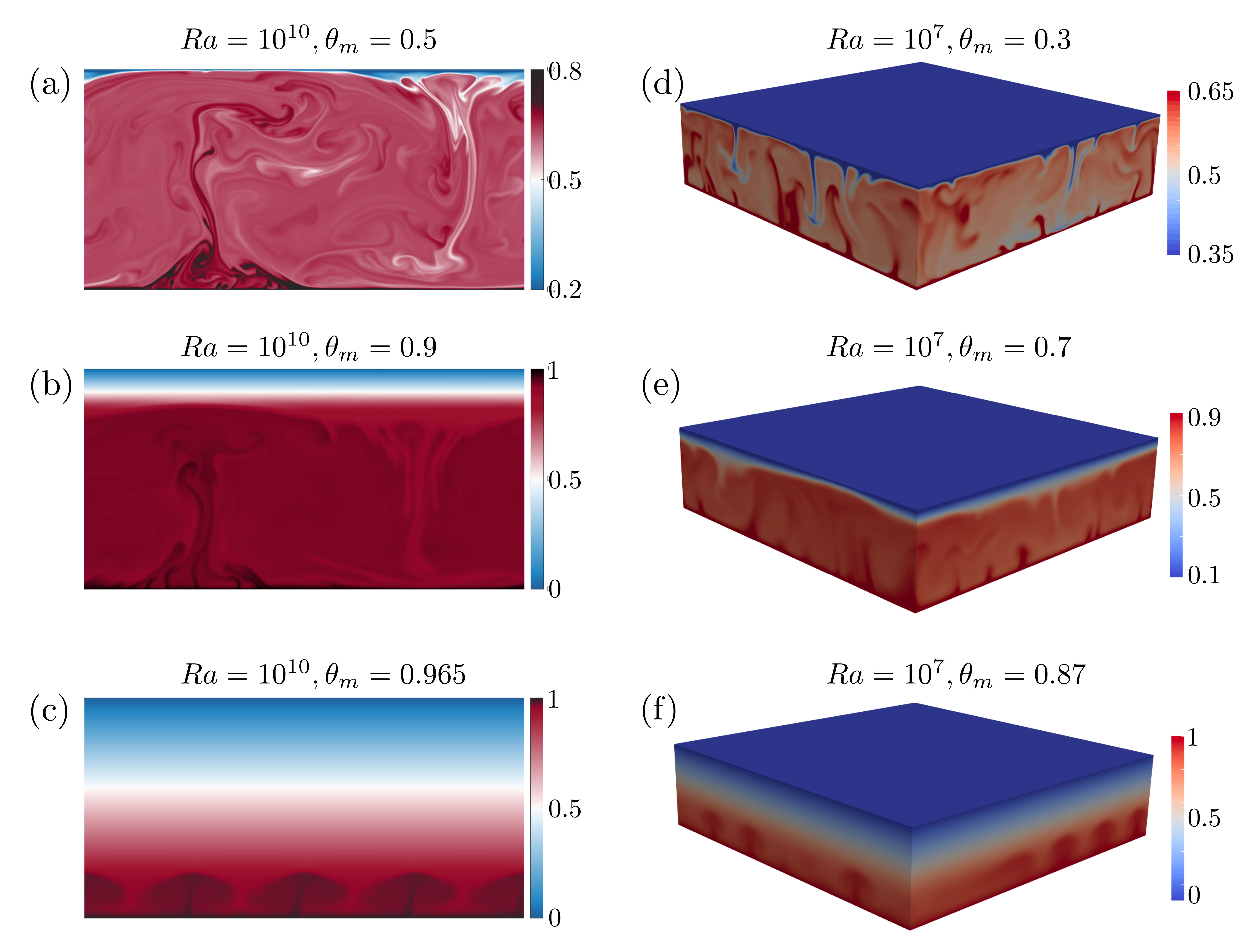}
   	\end{overpic}
\caption{
Instantaneous temperature fields for different $\theta_m$ in 
(a--c) 2D DNS for $Ra=10^{10}$ and $\Gamma=2$ and 
(d--f) 3D DNS for $Ra=10^{7}$ and $\Gamma=4$:
(a) $\theta_m=0.5$, 
(b) $\theta_m=0.9$, 
(c) $\theta_m=0.965$, 
(d) $\theta_m=0.3$, 
(e) $\theta_m=0.7$, and 
(f) $\theta_m=0.87$. 
}\label{flow}
\end{figure*}

The flow dynamics is governed by the incompressibility condition $\nabla\cdot\boldsymbol{u} = 0$ and the Navier-Stokes equation with (nonlinear, see above) 
buoyancy forcing
and the advection diffusion equation, 
\begin{eqnarray}
{\partial \boldsymbol{u}}/{\partial t} + \boldsymbol{u}\cdot\nabla\boldsymbol{u} &=& \sqrt{
{Pr}/{Ra}}\nabla^2\boldsymbol{u} -\nabla p + |\theta-\theta_m|^q{\vec{\boldsymbol{e}}_z}, \label{eq02}\\
{\partial \theta}/{\partial t} + \boldsymbol{u}\cdot\nabla \theta  &=& 
{1}/{\sqrt{Ra\,Pr}}\nabla^2\theta,\label{eq05}
\end{eqnarray}
\noindent where $\boldsymbol{u}=(u,w)$, $\theta$, and $p$ are the velocity, temperature, and  pressure, respectively. For non-dimensionalization, we choose $H$ and $U_f={(g\alpha \Delta ^qH)}^{1/2}$ as the reference length and velocity. The reference time is free-fall time $t_{f}=H/U_f$. Temperature is nondimensionalized as $\theta=(T-T_{t})/\Delta$. 
We consider periodic boundary conditions (BCs) in the horizontal direction  
and no-slip isothermal BCs at the top and bottom boundaries.

We perform 2D DNS in a broad $Ra$-range, $10^7\le Ra\le 10^{10}$, with fixed aspect ratio $\Gamma = 2$.
$Pr$ is fixed to $11.57$ in all simulations, which corresponds to the value for water at $4^\circ \rm{C}$.
In addition, we perform 3D DNS for $Ra=10^7$ and $\Gamma=4$. 
The governing equations are solved with the second-order finite-difference code  AFiD \citep{verzicco1996finite,van2015pencil}, which has already been extensively used to study RBC, see, e.g., \cite{wang2020multiple,wang2020zonal}. 
The used staggered grids satisfy the resolution requirements for DNS \citep{shishkina2010boundary}. 
Details on the simulations can be found in the Supplementary materials.

Fig.~\ref{flow} gives the reader an idea of the 
 flow organization.
  Figs.~\ref{flow}(a--c)
show 2D instantaneous temperature fields for different $\theta_m$ and $Ra=10^{10}$. 
For $\theta_m>0$, only the lower layer in the cell can be convectively unstable.
For not too large $\theta_m$, convection penetrates from the lower layer
into the upper, gravitationally stable, layer, and this increases the bulk temperature. 
For $\theta_m=0.5$ (Fig.~\ref{flow}~a), the convective flow occupies 
almost the whole domain, while 
for $\theta_m=0.9$ (Fig.~\ref{flow}~b) a stably stratified layer forms near the top plate. 
For even larger $\theta_m=0.965$ (Fig.~\ref{flow}~c), the flow stratification takes place nearly in the whole cell.
A similar change of the global flow structure with increasing $\theta_m$ is observed in the 3D DNS (Figs.~\ref{flow}~d--f).

The intrusion of penetrating convection from the convectively unstable regions to convectively stable ones
 resembles  those observed in internally heated convection \cite{goluskin2016penetrative,goluskin2016internally,wang2020ihc} and horizontal convection \cite{shishkina2016heat, shishkina2016prandtl, reiter2020classical}, as well as in counter-rotating Taylor--Couette flows, where stratified angular velocity zones are located near the outer cylinder \citep{van2011torque,ostilla2013optimal,ostilla2014exploring}.

\begin{figure*}
\centering
\begin{overpic}[width=1\textwidth]{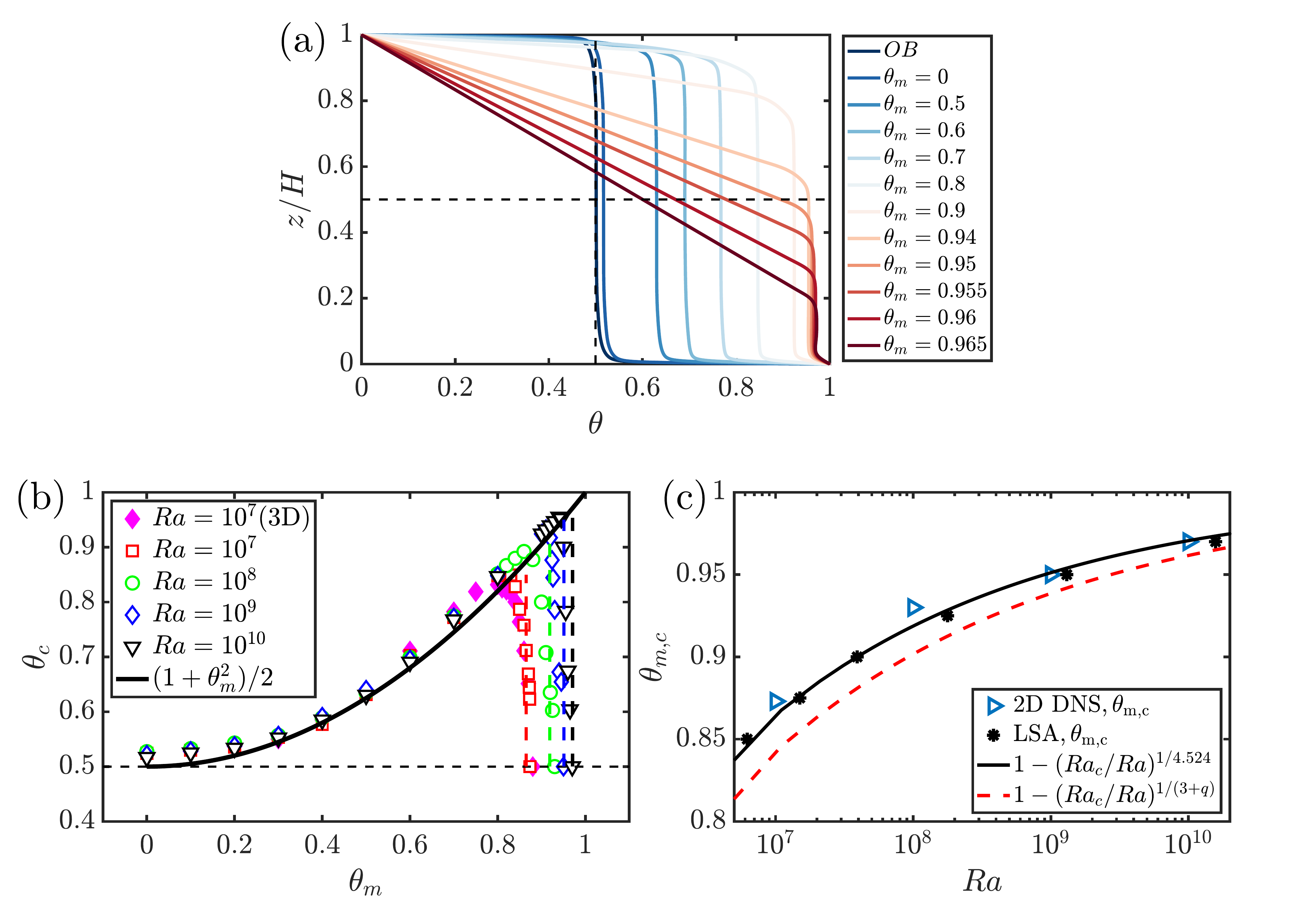}
\end{overpic}
\caption{
(a) Time- and horizontally averaged temperature profiles for different $\theta_m$, as obtained in the 2D DNS  for $Ra=10^{10}$ and $\Gamma=2$. 
(b) Time- and horizontally averaged temperature at mid-height, $\theta_c$, as function of $\theta_m$,
in the 2D (open symbols) and 3D (closed symbols) cases for different $Ra$.
The solid line shows the theoretical prediction $\theta_c=(1+\theta_m^2)/2$, which is universal for all $Ra$. 
Predictions of the linear stability analysis for the critical $\theta_m=\theta_{m,c}$, where $\theta_c$ drops back to 1/2, depend on $Ra$ and are shown with the dashed vertical lines of the colors that correspond to the DNS data.
(c) The critical $\theta_{m,c}$ versus $Ra$, as obtained from the 2D DNS (open symbols), linear stability analysis (closed symbols and their fit shown with the solid line) and from the theoretical model $\theta_{m,c}=1-(Ra_c/Ra)^{1/(3+q)}$, with $Ra_c=1171$ and $q=1.895$ (dashed line).}\label{tem_prof}
\end{figure*}

The time- and area-averaged temperature profiles and their dependences on $\theta_m$ for $Ra=10^{10}$ are presented in Fig.~\ref{tem_prof}(a).
It shows that for $\theta_m=0$, the profile is similar to that in the OB case, with the mean central temperature $\theta_c$
 being close to the arithmetic mean of the top and bottom temperatures 
 ($\theta_c=1/2$).
One can see that with increasing $\theta_m$, the temperature in the thermally unstable, lower zone gradually increases.
However, the mean temperature $\theta_c$ at mid-height 
behaves non-monotonically.
With increasing $\theta_m$, it first increases as long as the half height location ($z/H=0.5$) lies in the thermally unstable region, but then it starts to decrease back to the arithmetic mean of the top and bottom temperatures, once the mid-height location is swallowed up by the upper stably stratified region.

Fig.~\ref{tem_prof}(b) further shows the 
mean central  temperature $\theta_c$ as function of $\theta_m$, for both the 2D and the 3D cases. 
We note that the collapse of the  3D data with the 2D data  confirms 
 the similarity between 2D and 3D RBC for large $Pr$ \citep{van2013comparison}.
One can observe two different regimes: in regime I, for low density inversion parameter $0\le \theta_m \ll 1$,
the central temperature $\theta_c$ monotonically increases with increasing $\theta_m$, and all the data 
collapse  onto a single curve, independently of $Ra$. However, in  regime II, for $\theta_m$ close to 1, with increasing $\theta_m$
the central temperature $\theta_c$ extremely sharply drops  to the value 1/2.  
We denote the value of $\theta_m$, at which  $\theta_c$ reaches 1/2
in regime II, as $\theta_{m,c}$,
i.e., $\theta_c(\theta=\theta_{m,c})=1/2$.
It is clear that $\theta_{m,c}$ increases with increasing $Ra$ and approaches 1 for $Ra\rightarrow\infty$, implying that for strong enough thermal 
 driving the whole cell is filled with penetrative turbulence, even for large density inversion parameter close to 1.

We now set out to theoretically explain the universal dependence of  the central temperature $\theta_c$ 
on the control parameter $\theta_m$ in regime I.
 First we notice that for $\theta_m=0$, the temperature of the fluid is larger than $T_m$ through the entire convection cell and therefore the situation is similar to the OB case. Thus, the central temperature can be well approximated by the arithmetic mean of the top and bottom temperatures, i.e., $\theta_c=1/2$.
With increasing $\theta_m$, the height, at which the time- and area-averaged temperature equals $T_m$, gradually decreases from the top towards the bottom, and at a certain moment arrives at the mid-height, which indicates the end of  regime I.
Therefore, when $\theta_m\rightarrow1$ within regime I, the central temperature $\theta_c$ tends to $\theta_m$
and  $\lim_{\theta_m\rightarrow1}\frac{1-\theta_c}{1-\theta_m}=1$.
Applying L'H\^{o}pital's rule to this limit, we obtain that
$\lim_{\theta_m\rightarrow1}\frac{d\theta_c}{d\theta_m}=1$.
Thus, for $\theta_m\rightarrow1$, one can take $\theta_c=\theta_m$ and $d\theta_c/d\theta_m=1$. 
This together with $\theta_c=1/2$ at $\theta_m=0$ gives a simple polynomial approximation of the mean central temperature within the regime I,
\begin{eqnarray}
\theta_c=(1+\theta_m^2)/2. \label{thc03} 
\end{eqnarray}
Fig.~\ref{tem_prof}(b) demonstrates that the model (\ref{thc03}) accurately represents the 2D and 3D DNS data within regime I.

Next we will explain the abrupt drop  of the central temperature near $\theta_{m,c}$ as function of $Ra$.
 Therefore, 
 for any given $Ra$, we must
  find $\theta_m=\theta_{m,c}$, at which the central temperature $\theta_c$ drops back to 1/2 in regime II, see Fig.~\ref{tem_prof}(b).
As discussed above, the lower layer of the cell is convectively unstable for small $\theta_m$.
However, the situation changes dramatically for larger $\theta_m$, when the Rayleigh number $Ra_\ell$
 associated with this layer becomes as small as the critical Rayleigh number $Ra_c$ for the onset of convection in a domain with no-slip BCs at its lower surface and free-slip BCs at its top surface, at which the temperature is kept equal to $T_m$.
When at a certain $\theta_m=\theta_{m,c}$
 the value of $Ra_\ell$ becomes equal to $Ra_c$, the lower layer becomes convectively stable and stratified as the upper layer, and therefore the central temperature in the whole cell becomes equal to the arithmetic mean of the top and bottom temperatures of the cell, i.e., $\theta_c=1/2$.

The Rayleigh number $Ra_\ell$ associated with such stratified lower layer
 is related to $Ra$ as 
$Ra_\ell=(1-\theta_{m,c})^{3+q}Ra$.
This is due to the reduction of the height of the layer to $(1-\theta_{m,c})H$, compared to the height $H$ of the whole cell, and due to the reduction of the temperature difference between the boundaries of the layer to $(1-\theta_{m,c})\Delta$, compared to the temperature difference $\Delta$  at the cell plates.
Equating $Ra_\ell=Ra_c$, we obtain the prediction 
\begin{gather}
\theta_{m,c}=1-(Ra_c/Ra)^{1/(3+q)}. \label{thmc} 
\end{gather}
The critical $Ra_c$ for the onset of convection in an infinite layer of cold water with the top temperature $T_m$, no-slip BCs at the bottom and free-slip BCs at the top equals $Ra_c=1171$, as we
 calculated using linear stability analysis (LSA), see also Supplementary materials for more information.
The results produced by the model~(\ref{thmc}) are plotted as dashed line in Fig.~\ref{tem_prof}(c). 
The model prediction for  the behaviour of $\theta_{m,c}$ as function of $Ra$ is consistent with the  DNS  results (open symbols).

Even more accurate predictions of $\theta_{m,c}$ can be obtained with the LSA applied to a 2D convection cell  filled with cold water, for the cell aspect ratio $\Gamma=2$, 
and periodic BC at the side walls. Thus, for any given $\theta_m=\theta_{m,c}$, the LSA provides the critical Rayleigh number $Ra$ for the onset of convection; a relation between them is very precisely described by $\theta_{m,c}=1-(1171/Ra)^{1/4.524}$, see the Supplementary material. 
These predictions by the LSA are in a very good agreement with the DNS data, as it can be seen in 
Fig.~\ref{tem_prof}(c).

\begin{figure}
\centering
\begin{overpic}[width=1.0\textwidth]{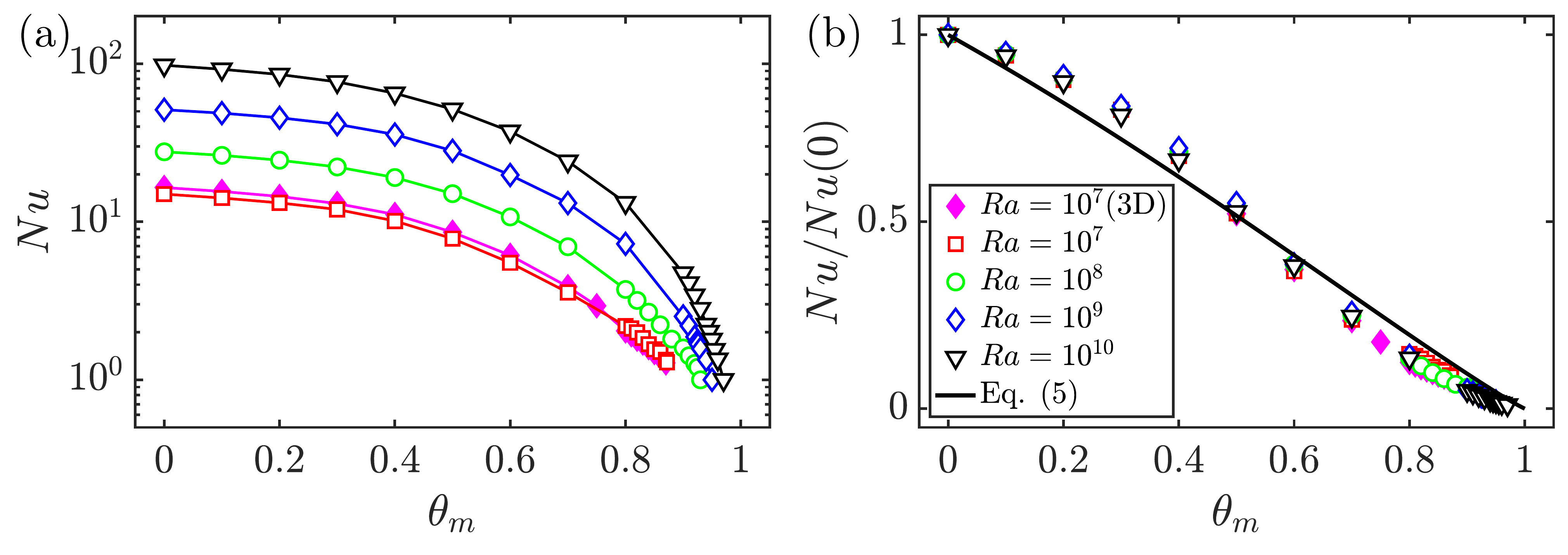}
\end{overpic}
\caption{
Nusselt numbers $Nu$ as function of the density inversion parameter $\theta_m$ for different $Ra$:
(a) Absolute values from the DNS (lines to guide the eye) and 
(b) normalized with the values at $\theta_m=0$.
The line  shows the theoretical model, Eq.~ (\ref{nu1}).
}\label{nu_re}
\end{figure}

We now focus on how the Nusselt number  $Nu$ depends
on the control parameter $\theta_m$ and develop a model for the reduced Nusselt number, $Nu(\theta_m)/Nu(\theta_m= 0)$, based on our knowledge on the heat transport properties in the OB case.
Fig.~\ref{nu_re}(a) shows that for all $Ra$, the absolute values of $Nu(\theta_m)$ monotonically decrease with 
growing $\theta_m$.
After normalization using $Nu(\theta_m=0)$, the data well collapse onto a single curve, as shown in Fig.~\ref{nu_re}(b).

How to account for  this universal relationship? 
For $\theta_m=0$, the whole RBC cell can be considered as an almost OB one, which is characterized by $Ra$ and the corresponding critical Rayleigh number $Ra_{c,0}$ for the onset of convection in this cell.
For $\theta_m>0$, only the lower layer can be treated as an OB cell.
The temperature at this layer's upper surface is $T_m$ and the corresponding Rayleigh number and critical Rayleigh number are $Ra_\ell$ and $Ra_{c}$, respectively.
From 
2D OB DNS with $Pr=10$ \cite{wang2020multiple} we know that in the considered $Ra$-range, $Nu$ scales as $(Ra/Ra_{c,0})^\gamma$ with $\gamma\approx0.27$.  Therefore
$Nu(\theta_m)/Nu(0)=(Ra_\ell/Ra_c)^\gamma(Ra_{c,0}/Ra)^\gamma=
(Ra_\ell/Ra)^\gamma(Ra_{c,0}/Ra_c)^\gamma$.
Due to the reduced height and the temperature drop in the lower layer compared to the whole RBC cell (by almost $(1-\theta_m)$, as a first approximation), the value of $Ra_\ell/Ra$ can be approximated as $(1-\theta_m)^4$.
The critical Rayleigh numbers for the onset of convection can be approximated as 
$Ra_{c}\sim [1+(1-\theta_m)^2/\Gamma^2]^2$ and
$Ra_{c,0}\sim [1+1/\Gamma^2]^2$
(see more information in the Supplementary material), which altogether lead to 
\begin{eqnarray}
\hskip-5mm\frac{Nu(\theta_m)}{Nu(0)}\approx\left[\frac{(1+\Gamma^2)(1-\theta_m)^2}{\Gamma^2+(1-\theta_m)^2}\right]^{2\gamma}=
\left[\frac{5(1-\theta_m)^2}{4+(1-\theta_m)^2}\right]^{0.54}.
\label{nu1}
\end{eqnarray}
This prediction is consistent with  the general trend of the normalized Nusselt number, as 
can be seen in Fig.~\ref{nu_re}(b).

\begin{figure}
\centering
\begin{overpic}[width=0.8\textwidth]{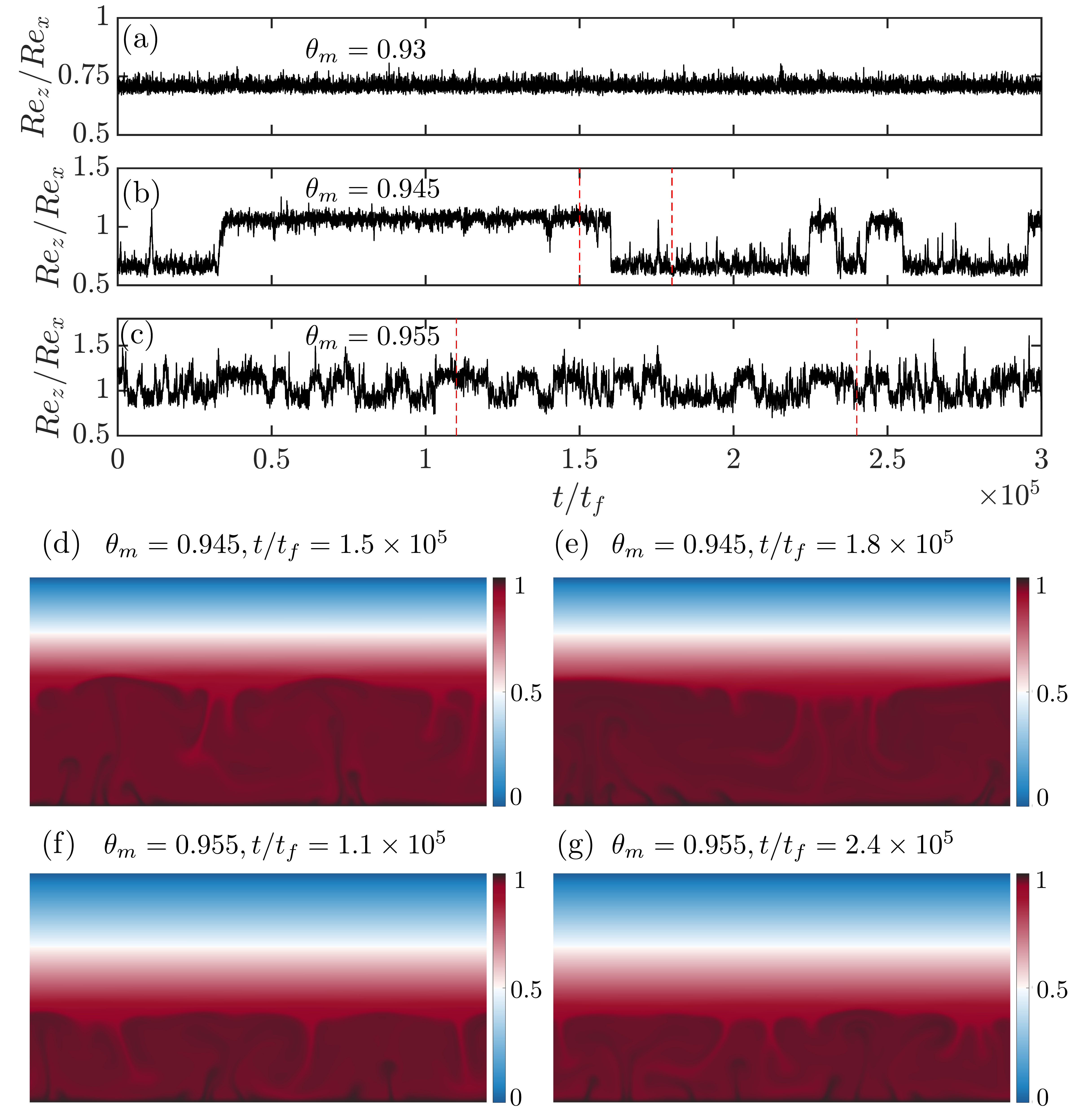}
\end{overpic}
\caption{
(a--c) Time evolution of the Reynolds number ratio $Re_z/Re_x$, as function of $\theta_m$, for $Ra=10^{10}$:
(a) $\theta_m=0.93$, 
(b) $\theta_m=0.945$,
(c) $\theta_m=0.955$.
(d--g) Instantaneous temperature fields for different $\theta_m$ at different times denoted by dashed lines in panels (b, c):
(d) $\theta_m=0.945$, $t/t_f=1.5\times10^5$;
(e) $\theta_m=0.945$, $t/t_f=1.8\times10^5$;
(f) $\theta_m=0.955$, $t/t_f=1.1\times10^5$ and
(g) $\theta_m=0.955$, $t/t_f=2.4\times10^5$.
}\label{re_ratio}
\end{figure}
Finally, we address the (rare) transitions between different flow states for large $\theta_m$.
It was found that the Reynolds number ratio $Re_z/Re_x$ can well differentiate different convection roll states \citep{wang2020multiple,wang2020zonal}. Here 
$Re_z = {\langle w^2\rangle_{V,t}^{1/2}}H/\nu$ 
is  the volume averaged vertical Reynolds number and 
$Re_x = {\langle u^2\rangle_{V,t}^{1/2}}H/\nu$ 
the horizontal one. The Reynolds number ratios $Re_z/Re_x$ for $Ra=10^{10}$ with large $\theta_m$ close to $\theta_m=1$ are displayed in Fig.~\ref{re_ratio}(a). The statistically stable two-roll state, as shown in Fig.~\ref{flow}(b), can be found for up to $\theta_m=0.93$. However, once $\theta_m$ increases to $\theta_m=0.945$, the two-roll state is not sustained all the time and rare transitions between two-roll and four-roll states occur, which can be seen in the rare jumps of the Reynolds number ratio $Re_z/Re_x$ in Fig.~\ref{re_ratio}(b). The temperature fields for the two-roll and four-roll states are shown in Figs.~\ref{re_ratio}(d) and \ref{re_ratio}(e), respectively. At even larger $\theta_m=0.955$, constant jumps between different states are observed, as displayed in Fig.~\ref{re_ratio}(c). The instantaneous temperature fields  at two instants are shown in Figs.~\ref{re_ratio}(f) and \ref{re_ratio}(g). It can be seen that the convection rolls are now not well organised. When $\theta_m$ reaches 0.97, the flow enters into the conduction state without any fluid motions.

Rare transitions between different turbulent states have been reported for many different flows \citep{sugiyama2010flow,van2011connecting,xie2018flow,wang2018multiple}. Similar phenomena are also important in many geophysical flows \citep{glatzmaier1995three,weeks1997}. Here we have thus found another example of such rare transition events in turbulent flows, which were not identified for OB cases \cite{wang2020multiple}. Similar to prior studies \citep{xie2018flow,wang2018multiple}, such phenomenon can be viewed as flow mode competition between different states: for not very large $\theta_m$, e.g., $\theta_m=0.93$, the effective aspect ratio of the convection region is close to $\Gamma=2$, therefore, it only supports a two-roll state. At large $\theta_m$ close to $\theta_m=1$, the effective aspect ratio becomes much larger than 1, and therefore it can support more convection rolls \citep{wang2020multiple}. In between, the different convection roll states may have comparable strength, and the competition of these states leads to the rare transitions between them.

In conclusion, based on our DNS of turbulent penetrative 
RBC in cold water, where the density anomaly temperature is achieved in the cell's bulk, we have shown that 
 the main response parameters, namely the  mean central temperature $\theta_c$
 and the  normalized Nusselt number, are universally 
 determined by the density inversion parameter $\theta_m$.
For $\theta_m=0$, the convective system is similar to that in the Oberbeck-Boussinesq case.
However, for $\theta_m$ approaching 1, the convection vanishes at a certain $\theta_m=\theta_{m,c}$ as soon as the Rayleigh number of the lower layer achieves the critical Rayleigh number $Ra_c$ for the onset of convection. 
We have demonstrated that $Ra_c$ and hence $\theta_{m,c}$ can be excellently predicted by a linear stability analysis.
We have further shown that in the range $0\le\theta_m \lesssim \theta_{m,c}$, the central temperature $\theta_c(\theta_m)$ 
is independent of $Ra$ and can be excellently predicted by our model.
Furthermore, we proposed a theoretical model for the normalized heat transport.

Our theoretical results on the similarity of the bulk temperature and normalized heat transport in penetrative thermal convection and the observation of the rare transitions between the flow states are supported by the DNS data. 
They are relevant and important for numerous geophysical, astrophysical and engineering applications. Our approach 
of theoretically determining the response parameters of penetrative convective turbulence in cold water
around $4^\circ$  may be transferred to other cases of penetrative convective turbulence.


\textit{Acknowledgement:} 
R. Verzicco is gratefully acknowledged for continuous collaborations and discussions. We also acknowledge the Twente Max-Planck Center and the Deutsche Forschungsgemeinschaft (Priority Programme SPP 1881 "Turbulent Superstructures" and grant Sh405/10).The simulations were carried out on the national e-infrastructure of SURFsara, a subsidiary of SURF cooperation, the collaborative ICT organization for Dutch education and research. Q.W. acknowledges financial support from the China Scholarship Council (CSC) and the Natural Science Foundation of China (NSFC) under grant no. 11621202.

\bibliography{reference}

\end{document}